\documentclass[letterpaper,prl,aps,twocolumn,floats,superscriptaddress, showpacs]{revtex4}
\usepackage{graphicx}
\usepackage{amsmath}
\newcommand{\ds}{\displaystyle}

\newcommand{\s}{\sigma}

\def\bea{\begin{eqnarray}}
\def\eea{\end{eqnarray}}
\def\be{\begin{equation}}
\def\ee{\end{equation}}

\begin{document}
\title{Scaling and universality in the 2D Ising model with 
a magnetic field}
\author{Vladimir V. Mangazeev}
\email[Corresponding author: ]{vvm105@physics.anu.edu.au}
\author{Michael {Yu}. Dudalev}
\affiliation{Department of Theoretical Physics, Research School of Physics and Engineering, Australian National University, Canberra, ACT 0200, Australia.}
\author{Vladimir V. Bazhanov}
\author{Murray T. Batchelor}
\affiliation{Department of Theoretical Physics, Research School of Physics and Engineering, Australian National University, Canberra, ACT 0200, Australia.}
\affiliation{Mathematical Sciences Institute, Australian National University, Canberra, ACT 0200, Australia.}
\begin{abstract}
The scaling function of the 2D Ising model in a magnetic field on 
the square and triangular lattices is obtained numerically via
Baxter's variational corner transfer matrix approach. 
The use of the Aharony-Fisher 
non-linear scaling variables allowed us to perform calculations
sufficiently away from the critical point to obtain very high
precision data, which convincingly confirm all predictions of the scaling
and universality hypotheses. 
The results are in excellent agreement with the field theory
calculations of Fonseca and Zamolodchikov as well as with many
previously known exact and numerical results for the 2D Ising model.
This includes excellent agreement with the classic 
analytic results for the magnetic
susceptibility by Barouch, McCoy, Tracy and Wu, recently
enhanced by Orrick, Nickel, Guttmann and Perk. 
\end{abstract}
\maketitle

The principles of scaling and universality (see, e.g.,
\cite{Car96}) play important roles in the theory of phase
transition and critical phenomena. The scaling assumption asserts that
observable quantities exhibit power law singularities in the variable 
$\Delta T=T-T_c$ in the vicinity of the critical temperature $T_c$, with  
coefficients being functions of certain dimensionless combinations of
available parameters, e.g., the magnetic field $H$ and $\Delta T$.
The universality hypothesis states that the leading singular part of
the free energy is a universal scaling function which is same for all
systems in a given ``universality class''.  In two dimensions classes
of universal critical behaviour are well understood --- they are
classified by conformal field theory (CFT) \cite{BPZ84}. 
The latter provides exact
solutions for 2D systems at the critical point and, in particular,
gives exact values of the scaling dimensions. However, the calculation
of scaling functions describing off-critical behaviour is a hard
problem which does not have any exact analytical solutions. From the
field theory point of view this requires solving a CFT perturbed by at
least two relevant operators. Direct numerical calculations and
simulations in lattice models also face serious difficulties due to
multiple long range correlations and the resulting poor convergence near the
critical point.

It appears that despite numerous analytical and numerical results
(cited below), the full picture of scaling and universality has never
been convincingly demonstrated through numerical
calculations in lattice models.  Our aim is to do this. Here we
consider the planar nearest-neighbour Ising model on the 
regular square and triangular lattices, which has already played a
prominent role in the development of the theory of phase transition
and critical phenomena
\cite{O44,Y52,McCoyWu,Baxterbook,AF83,BPZ84,Zam89a}.  Its
partition function reads 
\be 
{Z}=\sum_{\sigma}\exp \Big\{\,\beta
\sum_{\langle ij\rangle} \sigma_i \sigma_j +{H}\sum_i
\sigma_i\,\Big\}\ , \quad \sigma_i=\pm1,\label{Z-def} 
\ee 
where the
first sum in the exponent is taken over all edges, the second over all
sites and the outer sum over all spin configurations $\{\s\}$ of the
lattice. The constants $H$ and $\beta$ denote the (suitably
normalized) magnetic field and inverse temperature.  The specific free
energy, magnetization and magnetic susceptibility are defined as 
\be
F=-\lim_{N\to\infty}\frac{1}{N}\log Z,\quad M=-\frac{\partial
  F}{\partial H},\quad \chi=-\frac{\partial^2 F}{\partial
  H^2}\ ,\label{F-def} 
\ee 
where $N$ is the number of lattice
sites. The model exhibits a second order phase transition at $H=0$ and
$\beta=\beta_c$, where 
\be \beta_c^{(s)}={\textstyle{\frac{1}{2}}}
\log(1+\sqrt{2}),\qquad \beta_c^{(t)}= {\textstyle{\frac{1}{4}}}\log
3, 
\ee 
for the square \cite{O44} and triangular \cite{Newell1950}
lattices, respectively.

The scaling and universality 
hypotheses predict that the leading singular part,
 $F_{sing}(\Delta\beta,H)$, of the free energy in the vicinity of
 the critical point, $\Delta\beta=\beta-\beta_c\sim 0$, $H\sim0$, 
can be expressed through a universal 
function ${\mathcal F}(m,h)$,
\be
F_{sing}(\Delta\beta,H)={\mathcal F}(m(\Delta\beta,H),h(\Delta\beta,H)), 
\label{fsing}
\ee 
where $\Delta\beta$ and $H$ enter the rhs only through
non-linear scaling variables \cite{AF80},   
\bea
m&=&m(\Delta\beta,H)=O(\Delta\beta) +O\left((\Delta\beta)^3\right)
+O(H^2)+\ldots,\nonumber\\[.2cm]
h&=&h(\Delta\beta,H)=O(H)+H\,O(\Delta\beta)+O(H^3)\ldots,
\label{mh-def}
\eea 
which are analytic functions of $\Delta\beta$ and $H$. 
The coefficients in these expansions  
depend on the details of the microscopic interaction (for instance
they are different for the square and triangular lattices),
but the function
${\mathcal F}(m,h)$  is the same for all models in the 2D Ising model
universality class.  It can be written as 
\be
{\mathcal F}(m,h)=\frac{m^2}{8\pi} \, \log
m^2+
h^{16/15} \,\Phi(\eta), 
\quad \eta=\frac{m}{h^{8/15}}\label{Phi-def}
\ee
where $\Phi(\eta)$ is a universal scaling function of 
a single variable $\eta$ (the scaling parameter), normalized such that 
\be
{\mathcal F}(m,0)=\frac{m^2}{8\pi}\log m^2 \label{F-norm}
\ee
The function ${\mathcal F}(m,h)$ has a concise interpretation in 
terms of 2D Euclidean quantum field theory. Namely, 
it coincides with the vacuum energy density of 
the ``Ising Field Theory'' (IFT) \cite{FZ03}. The latter is defined as a
model of perturbed conformal field theory with the action
\be
{\cal A}_{\rm IFT} = {\cal A}_{(c=1/2)} + \frac{m}{2\pi}\,\int\,
\epsilon(x)\,d^2 x +
h\,\int \,\sigma(x)\,d^2 x\,,\label{IFT}
\ee
where ${\cal A}_{(c=1/2)}$ stands for the action of the $c=1/2$ CFT 
of free massless Majorana fermions, $\sigma(x)$ and
$\epsilon(x)$ are primary fields of conformal dimensions $1/16$ 
and~$1/2$. The parameters $m$ and $h$ have 
the mass dimensions $1$ and $15/8$, respectively,
and the scaling parameter $\eta$ in \eqref{Phi-def} is dimensionless.

The scaling function \eqref{Phi-def} is of much interest as it controls 
all thermodynamic properties of the Ising model in the critical
domain. 
Although there are many exact 
results (obtained through exact solutions of \eqref{IFT} 
at $h=0$ and all $\tau$ 
\cite{O44, Y52, BMW, TM73, McCoy76}, and at $\tau=0$ 
and all $h$ \cite{Zam89a, Fateev94,DM95,WNS92,WPSN94,BNW,Smi97,BS97};
these data 
are collected in \cite{Delfino2004}) as 
well as much numerical data \cite{Ess68, Zin96, VicC,VicA,VicB, Rut99}
about this function, its complete analytic characterization is still lacking. 

Recently \cite{FZ03} the function \eqref{Phi-def}, particularly its
analytic properties, have been thoroughly
studied in the framework of the IFT \eqref{IFT}.  
The authors of \cite{FZ03} made extensive numerical
calculations of the scaling function $\Phi(\eta)$ 
for real and complex values of $\eta$.

\begin{figure}
\includegraphics[scale=1.0]{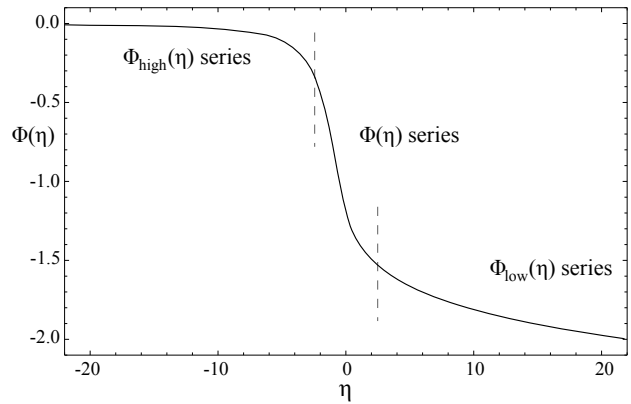}
\caption{The scaling function $\Phi(\eta)$ in the three regions,
  separated by the dashed lines at $\eta\simeq\pm 2.3$, 
can be approximated with high
precision (up to $10^{-6}$) by the series \eqref{Phi-high},
  \eqref{Phi-ser} and \eqref{Phi-low} with coefficients given in
Table 1. }\label{fig1}
\end{figure}

One of the motivations of our work was to confirm and extend 
the field theory results of \cite{FZ03} through {\em ab initio}
calculations, directly from the original lattice formulation 
\eqref{Z-def} of the Ising model. Here we give a brief summary of 
our results for the triangular and square lattices (the latter were
previously reported in \cite{MBBD08a}).
We used Baxter's variational
approach based on the corner transfer matrix method
\cite{BaxDim,BaxVar}.  
The main advantage of this approach over other numerical schemes (e.g., the
row-to-row transfer matrix method) is that it is formulated 
directly in the limit of an {\em infinite} lattice. 
Its accuracy depends on the magnitude of truncated eigenvalues of the corner
transfer matrix (which is at our control), rather than the size of the
lattice.  
The original Baxter approach \cite{BaxVar} was enhanced by an 
improved iteration scheme \cite{Nish}, known as the corner transfer
matrix renormalization group (CTMRG). The use of the non-linear
scaling variables \eqref{mh-def} allowed us to perform calculations
sufficiently away from the critical point with a reliable convergence
of the algorithm. 
The results for the scaling function $\Phi(\eta)$ are shown in
Fig.~\ref{fig1}. In total we have calculated about 10,000 data 
points for different values of the temperature and magnetic field on
the square lattice and about 5,000 for the triangular one. 
As seen from the picture all points collapse on a smooth
curve, shown by the solid line (as expected, the curve is the {\em same}
for the square and triangular lattices). 
The spread of the points at any fixed value $\eta$  
does not exceed $10^{-6}$ relative accuracy. This gives a convincing
demonstration of the scaling and universality in the 2D Ising model.
Furthermore, our numerical results for $\Phi(\eta)$ 
remarkably confirm the field theory calculations \cite{FZ03}, to
within all six significant digits presented therein.

For further reference we write down expansions of the function
$\Phi(\eta)$ for large values of $\eta$ on the real line
\begin{eqnarray}
\Phi_{low}(\eta) &=& \eta^2\sum_{k=1}^\infty\tilde{G}_k\,
\eta^{-{15k}/{8}},\qquad \eta \to +\infty\,,\label{Phi-low}\\[.3cm]
\Phi_{high}(\eta) &=& 
\eta^2 \sum_{k=1}^\infty G_{2k}\, |\eta|^{-30 k/8}, 
\quad \eta \to -\infty,\label{Phi-high}
\end{eqnarray}
and for small values of $\eta$, 
 \be
\Phi(\eta) = - {\frac{\eta^2}{8\pi}}\,\log\eta^2 + \sum_{k=0}^\infty
\Phi_k \eta^k.  \label{Phi-ser}
\ee
Some of the above expansion  coefficients are known exactly. 
The coefficient $\tilde G_1$ has a simple explicit expression \cite{McCoyWu};
the coefficients $G_2$ and $\tilde G_2$ have integral
expressions \cite{BMW,TM73} involving solutions of the Painlev\'e III
equation. They were numerically evaluated to very high precision
(50 digits) in \cite{ONGP}. 
The coefficients $\Phi_0,\Phi_1$ 
were analytically calculated in \cite{Fateev94} and \cite{FLZZ98},
respectively. The numerical value of $\Phi_1$ (which requires 
certain quadratures) was found in
\cite{MBBD08a}. The above values are quoted in the last column of
Table~1. 

In what follows we exclude the temperature variable $\beta$ in favour
of a new variable 
\be
\tau=\left\{\begin{array}{ll} (1-\sinh^2 2\beta)\,/\,(2\sinh 2\beta),&
\mbox{(sq. lat.)}\\[.2cm]
(e^{-\beta}-e^{\beta}\sinh2\beta)\,/\,(\sinh2\beta)^{1/2},&\mbox{(tr.
  lat.)}\end{array}\right.\label{tau-def}
\ee
which is vanishing for $\beta=\beta_c$ and positive for
$\beta<\beta_c$ (above the critical temperature). Another useful
variable 
\be
k^2=\left\{\begin{array}{ll} 16 \,e^{8\beta}/(e^{4\beta}-1)^4,&
\mbox{(sq. lat.)}\\[.2cm]
16\,e^{4\beta}/\big((e^{4\beta}-1)^3\,(e^{4\beta}+3)\big),
&\mbox{(tr.
  lat.)}\end{array}\right.\label{k-def}
\ee

The lattice free energy for $\tau,H\to0$, 
\be
F(\tau,H)=F_{sing}(\tau,H)+F_{reg}(\tau,H)+F_{sub}(\tau,H), 
\label{Ffull}
\ee
contains leading universal part \eqref{fsing}, 
regular terms $F_{reg}(\tau,H)$, which are analytic in $\tau$ and
$H$, and subleading singular terms $F_{sub}(\tau,H)$, 
which are non-analytic, but less singular than the first term in
\eqref{Ffull}.  
Therefore, to extract the universal scaling function 
from the lattice calculations one should be able to isolate and
subtract these extra terms. Moreover, one needs to
know the explicit form of the non-linear scaling variables
\eqref{mh-def}. In principle, all this information can be determined
entirely from numerical calculations (provided one assumes the values of
exponents of the subleading terms, predicted by the analysis
\cite{CPA02,ONGP} of
the CFT irrelevant operators, contributing to the free energy \eqref{Ffull}).
Much more accurate
results can be obtained if the numerical work is combined   
with known exact results. Namely, the zero-field free energy reads
\cite{O44,Newell1950}  
\be
\begin{array}{l}
F^{(s)}(\tau,0)= - \frac{1}{2} \log (4 \sinh 2\beta) - 
\frac{1}{8\pi^2}\iint\limits_0^{\ \ \ \ 2\pi}d\phi_1\, d\phi_2  \\[.2cm]
\qquad\qquad\qquad\log (2\sqrt{1+\tau^2}-\cos\phi_1 - \cos\phi_2 ),
\\[.2cm]
F^{(t)}(\tau,0) = 
- \frac{1}{2} \log (4 \sinh 2\beta) - 
\frac{1}{8\pi^2}\iint\limits_0^{\ \ \ \ 2\pi}d\phi_1\, d\phi_2  \\[.2cm]
\quad\log (3+\tau^2-\cos\phi_1 - \cos\phi_2 -\cos(\phi_1+\phi_2)),
\end{array}\label{Fint}
\end{equation}
where the superscipts (s) and (t) stand for the square and triangular
lattices, respectively.  
Write the non-linear variables  \eqref{mh-def} in the form,
\begin{eqnarray}
m(\tau,H)&=&-C_{\tau}\, \tau \ a(\tau)+H^2\,b(\tau)+O(H^4),\nonumber\\[.3cm]
h(\tau,H)&=&C_h H\,\Big[c(\tau)+H^2\,d(\tau)+O(H^4)\Big],\label{mh-def1}
\end{eqnarray}
where $a(0)=c(0)=1$, $h(\tau,H)=-h(\tau,-H)$ and write the
regular part in \eqref{Ffull} as, 
\be
F_{reg}(\tau,H)=A(\tau)+H^2\,B(\tau)+O(H^4)\ .\label{Freg}
\ee
As shown in \cite{ONGP}, 
the most singular subleading term, contributing to \eqref{Ffull} is of
the order of $\tau^{9/4}\,H^2\sim m^6$ for the square lattice 
and $\tau^{13/4}\, H^2\sim m^8$ for the
triangular lattice. 

Rewriting \eqref{Fint} in the form \eqref{Ffull} plus regular terms, 
one obtains 
\be
C^{(s)}_{\tau} = \sqrt{2}, \qquad C^{(t)}_{\tau} = 3^{-1/4}\,{\sqrt 2},
\ee
and 
\be
\begin{array}{l}
a^{(s)}(\tau) = { 1-\frac{3}{16}\tau^2+\frac{137}{1536}\tau^4+O(\tau^6)},
\\[.3cm]
a^{(t)}(\tau) =  1 - \frac{1}{24}\tau^2 + \frac{47}{10368}\tau^4
+O(\tau^{6}) \ .
\end{array}
\ee
The contribution to the regular part reads
%
\begin{equation}
\begin{array}{l}
A^{(s)}(\tau) = -\textstyle\frac{ 2{\mathcal G}}{ \pi}-\textstyle\frac{ \log 2}{ 2}+
\textstyle\frac{ 1}{ 2}\tau-\textstyle\frac{\ (1+5\log 2)}{\
  4\pi}\tau^2 
\\[.2cm] \qquad\qquad-
\textstyle\frac{ 1}{ 12}\tau^3+\textstyle\frac{ 5(1+6\log{2})}{ 64\pi}\tau^4
+O(\tau^5)\ ,\\[.4cm]
A^{(t)}(\tau) =
-\textstyle\frac{5}{2\pi}\text{Cl}_2\left(\frac{\pi}{3}\right)-\textstyle\frac{1}{4}\log\frac{4}{3}
+ \textstyle \frac{\tau}{3} 
- \textstyle\Big(\frac{2 + 3\log 12}{8\pi\sqrt 3} \\[.2cm] 
- \textstyle\frac{1}{36}\Big)\tau^2 -
\textstyle \frac{7}{648} \tau^3 \textstyle
 + \left(\frac{4+9\log 12}{288\pi\sqrt 3} - \frac{1}{324}\right)
 \tau^4 + O(\tau^5).
\end{array}
\end{equation}
Next, with the definition \eqref{k-def} the zero-field spontaneous
magnetization has the same expression for both lattices 
\be
M(\tau,0)=(1-k^2)^{1/8},\qquad \tau<0\ .
\ee 
Combining this with \eqref{F-def}, \eqref{Phi-def}, \eqref{Phi-low}  and 
\eqref{Ffull} one obtains 
\be
C^{(s)}_h  = -2^{3/16}/\tilde{G}_1, \qquad C^{(t)}_h  = -2^{5/16} 3^{-3/32}/\tilde{G}_1,
\ee
and 
\be\begin{array}{l}
c^{(s)}(\tau)=\ds 1+\frac{\tau}{4}+
\frac{15\tau^2}{128}-\frac{9\tau^3}{512}-\frac{4333\tau^4}{98304}+O(\tau^5),
\\[.3cm]
c^{(t)}(\tau)=\ds 1+\frac{\tau}{6} +\frac{5\tau^2}{96}+\frac{\tau^3}{576}-
\frac{727\tau^4}{165888} + O(\tau^5).
\end{array}
\ee
Finally, consider the zero-field susceptibility. 
The second field derivative of \eqref{Ffull} at $H=0$ gives
\be
\begin{split}
 \chi(&\tau)= - \frac{2\,G\,{C_h^2 \,c(\tau)^2}}{(\sqrt2 \,
|\tau|\,a(\tau))^{7/4}}   -\frac{\partial^2 F_{sub}}{\partial
H^2}\Big\vert_{H=0}
\\
& - 2B(\tau) +\frac{\tau\,a(\tau)\,b(\tau)}{\sqrt{2}\,\pi}
\big(1+\log (2\tau^2 a(\tau))\big) , \label{sus}
\end{split}
\ee
where  $G=G_2$ for $\tau>0$ and $G=\tilde{G}_2$ for $\tau<0$.
No simple closed form expression for the zero-field susceptibility
$\chi(\tau)$ is known. However, the authors of \cite{ONGP} obtained
remarkable asymptotic expansions of $\chi(\tau)$ for the square
lattice for small
$\tau$ to within $O(\tau^{14})$ terms with high-precision numerical 
coefficients. Using their results in \eqref{sus}, one obtains 
\be
\begin{split}
B^{(s)}(\tau)&=0.0520666225469+0.0769120341893\, \tau\\
& +0.0360200462309\,\tau^2+O(\tau^3)\label{Btau} ,
\end{split}
\ee
and 
\be
b^{(s)}(\tau)=\mu^{(s)}_h\,\Big(1+\frac{\tau}{2}+
 O(\tau^2)\Big),\quad \mu^{(s)}_h=0.071868670814
\ee
No similar expansion for $\tau \sim 0$ is available for the triangular
lattice. We used our data for $\tau=0$ to estimate
\be
B^{(t)}(\tau)=0.0247805582(2)+O(\tau),\quad \mu_h=-0.010475(1)
\ee
and the coefficient $d(\tau)=e_h+O(\tau)$ in \eqref{mh-def1} 
\be
e^{(s)}_h=-0.00728(30),\quad
e^{(t)}_h=+0.00129(1),
\ee
which is in agreement with $e^{(s)}_h=-0.00727(15)$ from \cite{VicC}. 

The above expressions were used to analyze our extensive numerical data and
extract the necessary information to obtain the universal scaling function. The
results are summarized in Table~1. For convenience of
comparison we quoted the field theory results from \cite{FZ03}.
Earlier exact and numerical results for the
same quantities are also quoted (whenever available). 

To conclude, we have implemented Baxter's variational corner transfer
matrix approach to obtain the universal scaling function for the Ising
model in a magnetic field on the square
and triangular lattice, as shown in Fig.~1 and
Table~1.
The numerical data is seen to be in remarkable agreement with the field theory
results obtained by Fonseca and Zamolodchikov \cite{FZ03}.
We also report a remarkable agreement (11 to 14 digits) between 
our numerical values for $\tilde{G}_1$,  $G_2$ and 
$\tilde{G}_2$ and the classic exact results of Barouch, McCoy, Tracy and
Wu \cite{McCoyWu,BMW,TM73} and a similar
agreement between the values $\Phi_0$ and $\Phi_1$ and the exact predictions
\cite{Fateev94,FLZZ98} of Zamolodchikov's integrable $E_8$ field
theory \cite{Zam89a}. Interestingly, this $E_8$ symmetry has now been
observed in experiments on the transverse Ising chain \cite{Col10}. 

The authors thank H.~Au-Yang, R.J.~Baxter, G.~Delfino, M.E.~Fisher, 
A.J.~Guttmann, 
S.L.~Lukyanov, S.B.~Rutkevich, 
C.A.~Tracy and
A.B.~Zamo\-lod\-chi\-kov for useful discussions and remarks and to
J.H.H.~Perk for also providing us with his unpublished high- and
low-temperature   
series for the triangular lattice Ising model. 
This work has been partially supported by the Australian Research Council.
   
\begin{widetext}

\small
\begin{table}[h]
\centering
\begin{tabular}{l|l|l|l|ll}
& \multicolumn{1}{|c|}{Tr. lat. CTM} 
& \multicolumn{1}{|c|}{Sq. lat. CTM} 
& \multicolumn{1}{|c}{IFT \cite{FZ03}} 
& \multicolumn{1}{|c}{References} \\
\hline
$\tilde G_1$\rule{0cm}{.4cm} & $-1.357838341706595(2)$   &$-1.3578383417066(1)$  & $-1.35783835$ & $  -1.357838341706595496... $ & \cite{McCoyWu}\\
$\tilde G_2$ & $-0.048953289720(2)$ &$ -0.048953289720(1)$ & $-0.0489589$ & $ -0.0489532897203... $ & \cite{BMW,TM73,ONGP} \\
$\tilde G_3$ & $\phantom{+} 0.0388639290(1)$ &$ \phantom{+} 0.038863932(3)$ & $\phantom{+} 0.0388954$& $\phantom{+}0.0387529$ \cite{MW78}; $\phantom{+}0.03893$&\cite{Rut99}\\
$\tilde G_4$ & $-0.068362121(1)$ & $-0.068362119(2)$ & $-0.0685060$& $ -0.0685535$ \cite{MW78}; $-0.0685(2)$ &\cite{Zin96} \\
$\tilde G_5$ & $\phantom{+}0.18388371(1)$ & $\phantom{+}0.18388370(1)$  & $\phantom{+}0.18453$ & \multicolumn{1}{|c}{---}\\
$\tilde G_6$ & $-0.659170(1)$ & $-0.6591714(1)$ & $-0.66215$& \multicolumn{1}{|c}{---}\\
$\tilde G_7$ & $\phantom{+}2.93763(2)$ & $\phantom{+}2.937665(3)$& $\phantom{+}2.952$&\multicolumn{1}{|c}{---}\\
$\tilde G_8$ & $-15.57(2)$ & $-15.61(1)$ & $-15.69$&\multicolumn{1}{|c}{---}\\
\hline
$G_2$ \rule{0cm}{.4cm}& $ -1.84522807823(1)$  &$-1.8452280782328(2)$  & $-1.8452283$& $ -1.845228078232838...$ &\cite{BMW,TM73,ONGP} \\
$G_4$ & $\phantom{+}8.3337117508(1)$ & $\phantom{+}8.333711750(5)$ & $\phantom{+}8.33410$& $\phantom{+}8.33370(1)$&\cite{VicC}\\
$G_6$ & $-95.16897(3)$ & $-95.16896(1)$ & $-95.1884$& $-95.1689(4) $&\cite{VicC}\\
$G_8$ &  $\phantom{+}1457.8(2)$ & $\phantom{+}1457.62(3)$ & $\phantom{+}1458.21$& $\phantom{+}1457.55(11)$&\cite{VicC}\\
\hline
 $\Phi_0$ \rule{0cm}{.4cm}& $-1.197733383797993(1)$ & $ -1.197733383797993(1)$  
 & $-1.1977320$ & $-1.19773338379799339...$  &\cite{Fateev94}\\
$\Phi_1$ & $-0.3188101248906(1)$ & $-0.318810124891(1)  $ 
 & $-0.3188192$ & $-0.31881012489061...$ & \cite{FLZZ98,MBBD08a}\\ 
$\Phi_2$& $\phantom{+}0.1108861966832(3) $ & $\phantom{+}0.110886196683(2)$ 
 & $\phantom{+}0.1108915$ & \multicolumn{1}{|c}{---}\\ 
$\Phi_3$& $\phantom{+}0.01642689465(1)$ & $\phantom{+}0.01642689465(2)$
 & $\phantom{+}0.0164252$ & \multicolumn{1}{|c}{---}\\
$\Phi_4$& $-2.6399783(1)\times10^{-4}$ & $-2.639978(1)\times10^{-4}$ 
& $-2.64\times10^{-4}$ & \multicolumn{1}{|c}{---}\\
$\Phi_5$& $-5.140526(1)\times10^{-4}$ & $-5.140526(1)\times10^{-4}$ 
 & $-5.14\times10^{-4}$ & \multicolumn{1}{|c}{---}\\
$\Phi_6$& $\phantom{+}2.08866(1)\times10^{-4}$ & $\phantom{+}2.08865(1)\times10^{-4}$ 
 & $\phantom{+}2.09\times10^{-4}$ & \multicolumn{1}{|c}{---}\\
$\Phi_7$& $-4.481969(2)\times 10^{-5}$ & $-4.4819(1)\times 10^{-5}$
 & $-4.48\times10^{-5}$& \multicolumn{1}{|c}{---}\\
$\Phi_8$& $\phantom{+}3.194(1)\times10^{-7}$ & \multicolumn{1}{|c|}{---}
 & $\phantom{+}3.16\times10^{-7}$&\multicolumn{1}{|c}{---}\\
$\Phi_9$& $\phantom{+}4.313(1)\times10^{-6}$ & \multicolumn{1}{|c|}{---}
 & $\phantom{+}4.31\times10^{-6}$&\multicolumn{1}{|c}{---}\\
$\Phi_{10}$& $-1.987(2)\times10^{-6}$ & \multicolumn{1}{|c|}{---}
 & $-1.99\times10^{-6}$&\multicolumn{1}{|c}{---}\\
$\Phi_{11}$ & $\phantom{+}4.32(1)\times10^{-7}$ & \multicolumn{1}{|c|}{---}
 & \multicolumn{1}{|c|}{---} & \multicolumn{1}{|c}{---}
\end{tabular}
\caption{Numerical values of the coefficients 
$G_n$, $\tilde G_n$, $\Phi_{n}$} 
\label{tablef}
\end{table}

\end{widetext}
\newcommand\oneletter[1]{#1}

\end{document}